\documentclass[12pt]{article}
\usepackage{amsmath}
\usepackage{amssymb}
\usepackage{float}
\usepackage{graphicx}
\usepackage{tabularx,ragged2e,booktabs,caption}
\usepackage{caption}
\usepackage{subcaption}
\usepackage{hyperref}
\usepackage{slashed}
\usepackage{multicol}
\usepackage[margin=1.0in,footskip=0.25in]{geometry}

\begin{document}
\begin{center}
{\bf Evolution of Dark Energy Perturbations for Slotheon Field and Power Spectrum}
\end{center}
\vspace{1cm}
\begin{center}
 {\bf Upala Mukhopadhyay $^1$, Debasish Majumdar $^2$}
\end{center}
\vskip 0.5mm
\begin{center}
$^{1,2}$Astroparticle Physics and Cosmology Division  \\
Saha Institute of Nuclear Physics, HBNI  \\
1/AF Bidhannagar, Kolkata 700064, India  \\
\end{center}
\begin{center}
{\bf Debabrata Adak$^3$}
\end{center}
\begin{center}
$^3$Department of Physics, Government General Degree College,\\ Singur, Hooghly 712409, West Bengal, India.
\end{center}
\bigskip
\bigskip
\bigskip
\begin{center}
{\bf Abstract} 
\end{center}
{\small
Within the framework of modified gravity model namely Slotheon model, inspired by the theory of extra dimensions, we explore the behaviour of Dark Energy and the perturbations thereof. The Dark Energy and matter perturbations equations are then derived and solved numerically by defining certain dimensionless variables and properly chosen initial conditions. The results are compared with those for standard quintessence model and $\Lambda$CDM model. The matter power spectrum is obtained and also compared with that for $\Lambda$CDM model. It appears that Dark Energy in Slotheon model is more akin to that for $\Lambda$CDM model than the standard quintessence model.}

\
\
\bigskip
\newpage
\tableofcontents
\newpage
\section{Introduction}

In modern cosmology one of the most challenging problems is to explain the late time acceleration of the Universe. In 1998 it was first discovered by using Supernova Type Ia observation \cite{SN1},\cite{SN2} that the Universe is not only expanding,
it is accelerating. Since then many observations like Cosmic Microwave Background Radiation observation \cite{CMB1}-\cite{CMB2}, baryon acoustic oscillations measurement in galaxy power spectrum \cite{BAO1},\cite{BAO2}, large scale structure observation \cite{LSS1}-\cite{LSS2} have also supported this phenomenon of late time acceleration of the Universe. In order to describe this accelerated expansion cosmologists have introduced the concept of Dark Energy, a component with negative pressure, which contains $\sim$ 68.5$\text{\%}$ of the total energy density of the Universe at the present epoch.

As mentioned, the pressure acts opposite to that of gravity so as to accelerate the Universe in opposition to the tendency of an eventual gravitational collapse due to the mass present in the Universe. Several theoretical models have since been proposed for explaining the origin and nature of the mysterious Dark Energy and consequent late time acceleration. There are attempts to introduce this Dark Energy by the cosmological constants in the framework of Einstein's equations and the Friedmann equations
that follows for FRW cosmology \cite{sami}-\cite{lambda2}. However from the point of view of particle physics the cosmological constant naturally arises as an energy density of the vacuum, but if $\Lambda$ originates from a vacuum energy density, then
one has to make a fine tuning of the theoretical result of order $10^{121}$ \cite{sami}, \cite{wein}, \cite{lamprob}. Cosmological constant $\Lambda$ is also associated with another theoretical problem, namely cosmic coincidence problem \cite{coincidence}. These have motivated to study other Dark Energy models to explain late time acceleration of the Universe. There are efforts where Dark Energy density is not considered to be a constant but of varying in nature and to explain such a time varying Dark Energy or quintessence Dark Energy \cite{Tsujikawa:2013fta}a scalar field in a scalar potential \cite{potential} that changes as time progresses have been invoked. Various scalar field Dark Energy models have been studied in detail in literature \cite{models}-\cite{model2}. Other attempts to explain this late time acceleration includes modification of Einstein's gravity in the large scale due to the presence of large extra dimensions \cite{DGPnote}. There are various proposals in the literature of higher dimensional models \cite{branemodel} to explain recent cosmic acceleration.

The Dark Energy and the inhomogeneity in the Dark Energy field are addressed by studying the Dark Energy perturbations \cite{SDEP},\cite{Gal}. This is related to the perturbations of space time \cite{dodleson}-\cite{a.das} as also the perturbations of the scalar field that may be considered to account for the Dark Energy. The study of cosmological perturbations is also important because the matter and other perturbations that are derived from a proposed theory have direct consequences in accounting for the matter power spectrum.

In the present work we investigate the late time acceleration by a scalar field namely Slotheon \cite{slotheon},\cite{slotheon_another} field inspired by extra dimensional models. A Slotheon field arises out of the Dvali, Gabadadze and Porrati (DGP) \cite{DGP} model related to brane world. At the limit when Planck mass $M_{\rm {pl}} \rightarrow \infty$ and $r_c \rightarrow \infty$, where $r_c$ is a cross over scale for transition from 4-dimension to 5-dimension (1 extra dimension), this theory in the Minkowski space time can be described by a scalar field \cite{Galshift},\cite{176}, where the field obeys a shift symmetry (Galileon shift) \cite{Galshift}. Here the strong coupling scale of the DGP model given by $(r_c^2/M_{\rm {pl}})^{1/3}$ \cite{Galshift} remains fixed. A suitable scalar field that describes this symmetry when extended to the curved space time \cite{galshift} is termed as Slotheon scalar field. We calculate in the present work the general relativistic
perturbations \cite{a.das},\cite{pert} for Slotheon field model and derive the analytical expression for fractional matter density perturbation, fractional density perturbation of the Slotheon field considered  and other relevant quantities. Evolution of these perturbations are also worked out and shown in detail in the present work. Finally the matter power spectrum for the Slotheon field is computed and shown.

This paper is organised as follows. In section 2, we discuss the background evolution of the Universe for Slotheon field model, in section 3 we talk about first order general relativistic perturbation equations and there solutions. Section 4 is devoted to the calculations of evolution of different perturbative quantities  with time considering the Slotheon field. In section 5 we investigate the effect of the Slotheon field on the matter power spectrum and finally in section 6 a summary and discussions are given.

\section{Background Evolution for Slotheon Field}

Slotheon field is a scalar field model inspired by a model in theories of extra dimensions, which is a class of modified gravity models and can be an alternative way to explain the late time acceleration of the Universe. As mentioned earlier the Slotheon model follows from the DGP model with one extra dimension. 

Writing down the DGP action in Minkowski space time in terms of a scalar field $\pi$ (called the Galileon field) and then writing the equations of motion up to the second order derivative of $\pi$ the theory is made free from the ghost degrees of freedom. The scalar field $\pi$, which is called the Galileon field obeys a shift symmetry
\begin{equation}
\partial_\mu \pi \rightarrow \partial_\mu \pi + c_\mu \,\,,
\end{equation}
termed as Galileon shift symmetry. The Galileon shift symmetry in the flat Minkowski space time can also be given by
$\pi \rightarrow \pi + a + b_\mu x^\mu$, where $a$ and $b_\mu$ are denoting a constant and a constant vector respectively.
The Slotheon field is obtained when the Galileon shift is extended to curved space time \cite{galshift}. Here we work with a Slotheon field which is described in what follows.
 
The Lagrangian $ \mathcal{L} = -\dfrac{1}{2} g^{\mu\nu} \pi_{;\mu} \pi_{;\nu} + \dfrac{G^{\mu\nu}}{2M^2} \pi_{;\mu} \pi_{;\nu}  $ ($\pi_{;\mu}$ denotes the covariant derivative of $\pi$, $G_{\mu\nu}$ is the Einstein's tensor while $g_{\mu\nu}$ is the metric and $M$ being an energy scale) remains invariant under this shift in curved space time. By adding standard Einstein-Hilbert term to the Lagrangian density $\mathcal{L}$ and a non trivial potential for $\pi$, one obtains a rich gravitational theory, with some interesting properties. The field $\pi$ moves slower in this theory than in the scalar field theory described by the Lagrangian $ \mathcal{L} = -\dfrac{1}{2} g^{\mu\nu} \pi_{;\mu} \pi_{;\nu} $. For this reason $\pi$ is called a Slotheon field. 

The Slotheon action is given as \cite{slotheon}
\begin{equation}
S = \int d^4x \sqrt{-g} \left[\dfrac{1}{2} \left(M_{{\rm pl}}^2 R - \left(g^{\mu\nu} - \dfrac{ G^{\mu\nu}}{M^2} \right)\pi_{;\mu} \pi_{;\nu}\right) - V({\pi})\right)+S_m \left[\psi_m; e^{2\beta\pi/M_{\rm pl}} g_{\mu\nu}\right]\,\,, \ \label{action}
\end{equation}
where $ M_{{\rm pl}}^2 = \dfrac{1}{8 \pi G} $ is the reduced Planck mass, $M$ is an energy scale and $S_m$ is the action for the matter field. In the above $R$ is the Ricci scalar, $\psi_m$ is the matter field that couples to the field $\pi$ with a dimensionless coupling constant $\beta$. It can be noted here that without the term $\dfrac{G^{\mu\nu}}{2M^2}\pi_{;\mu}\pi_{;\nu}$, the action of Eq. (\ref{action}) is similar to the action of a standard quintessence scalar field \cite{sami}. Variations of this action with respect to the metric and the field $\pi$ yield the following equations of motion respectively
\begin{eqnarray}
 M_{{\rm pl}}^2  G_{\mu\nu} & = &  T_{\mu\nu}^{(m)} + T_{\mu\nu}^{(\pi)}\,\,, \\
\Box \pi + \dfrac{1}{M^2}\left[\dfrac{R}{2} \Box \pi - R^{\mu\nu} \pi_{;\mu\nu}\right]-V^\prime(\pi)  & = &  -\dfrac{\beta}{M_{{\rm pl}}} T^{(m)}\,\,,
\end{eqnarray}
where $T_{\mu\nu}^{(m)}$, $T_{\mu\nu}^{(\pi)}$ are energy momentum tensors of dust like particles and the field respectively. In what follows we will not consider $T_{\mu\nu}^{(r)}$ (energy momentum tensor for radiation) in this work since the initial conditions adopted in the present work do not include the radiation dominated era as is described later. In this case
\begin{equation}
\begin{split}
T_{\mu\nu}^{(\pi)} &= \pi_{;\mu} \pi_{;\nu} -\dfrac{1}{2} g_{\mu\nu} (\nabla \pi)^2 - g_{\mu\nu} V(\pi) + \dfrac{1}{M^2}\left( \dfrac{1}{2} \pi_{;\mu} \pi_{;\nu} R -2 \pi_{;\alpha} \pi_({_{;\mu} R_{\nu}^{\alpha}})\right.\\
& \left. +\dfrac{1}{2}\pi_{;\alpha} \pi^{;\alpha}G_{\mu\nu} - \pi^{;\alpha} \pi^{;\beta} R_{\mu\alpha\nu\beta} -\pi_{;\alpha\mu} \pi_{;\nu}^{\alpha} + \pi_{;\mu\nu} \pi_{;\alpha}^{\alpha} + \dfrac{1}{2} g_{\mu\nu}\left(\pi_{;\alpha\beta} \pi^{;\alpha\beta} - (\pi_{;\alpha}^{\alpha})^2 + 2\pi_{;\alpha} \pi_{;\beta} R^{\alpha\beta}\right)\right)\,\,. \label{slotheon_T}
\end{split}
\end{equation}
In spatially flat Friedmann Robertson Walker (FRW) background with the assumption that the coupling constant $\beta=0$, the equations of motion take the form
\begin{equation}
3 M_{{\rm pl}}^2 H^2 = \rho_m  + \dfrac{\dot{\pi}^2}{2} + \dfrac{9 H^2 \dot{\pi}^2}{2M^2} + V(\pi)\,\,, \label{EE1}
\end{equation}
\begin{equation}
M_{{\rm pl}}^2 (2\dot{H} + 3H^2) = -\dfrac{\dot{\pi}^2}{2} +V(\pi) + (2\dot{H} + 3H^2) \dfrac{ \dot{\pi}^2}{2M^2} + \dfrac{2 H\dot{\pi}\ddot{\pi}}{M^2}\,\,, \label{EE2}
\end{equation}
\begin{equation}
0 =\ddot{\pi} + 3H\dot{\pi} + \dfrac{3 H^2}{M^2}\left(\ddot{\pi} + 3H\dot{\pi} + \dfrac{2\dot{H}\dot{\pi}}{H}\right) + V_{\pi}\,\,. \label{EE3}
\end{equation}
Here dot represents derivative w.r.t. time and double dot represents double derivative w.r.t. time. The derivative of $V(\pi)$ w.r.t $\pi$ is given as $V_\pi$. It can be noted that $\pi$ field is slower than a canonical scalar field and hence the name Slotheon. The slowing of the field is entirely due to gravitational interaction. In the present analysis we adopt an
exponential form of the potential $V(\pi)$ given by
\begin{equation}
V(\pi)=V_0 \rm exp \left(-\dfrac{\lambda \pi}{M_{\rm pl}}\right)\,\,, \label{potential}
\end{equation}
where $\lambda$ is a constant.
\section{Perturbations for the Slotheon Field}

In the present work the cosmological perturbations for both the Slotheon field and the matter are carried out in the longitudinal gauge or Newtonian gauge \cite{DE}. The scalar perturbed metric under this framework is given by \cite{dodleson},\cite{pert}
\begin{equation}
ds^2=-(1+2 \Phi) dt^2+a^2(t) (1+2 \Psi) \delta_{ij} dx^idx^j\,\,, \label{metric}
\end{equation}
where $a(t)$ is the scale factor, $\Phi$ is gravitational potential and $\Psi$ is the perturbation in the spatial curvature. The anisotropic stress is assumed to be zero and therefore $\Phi=-\Psi$ \cite{DE}. Under this circumstance the perturbed metric can be completely described by a single scalar variable $\Phi$. 
It is assumed that the baryonic matter, dark matter etc. can be described as perfect fluid so that the energy momentum tensor is written as 
\begin{equation}
T^\mu_\nu = (\rho + p)u^\mu u_\nu + p \delta^\mu_\nu\,\,, \label{energy_momentum}
\end{equation}
where $\rho$, $p$, $u_\mu$ are respectively the energy density, pressure density and four velocity of the fluid.
The perturbations in these and the Slotheon field $\pi$ are defined as
\begin{eqnarray}
\rho(t ,\overrightarrow{x}) & = & \bar{\rho}(t) + \delta \rho(t ,\overrightarrow{x})\,\,,\\
p(t ,\overrightarrow{x}) & = & \bar{p}(t) + \delta p(t ,\overrightarrow{x})\,\,,\\
u^\mu & = & \bar{u}^\mu + v^\mu\,\,, \label{velocity} \\
\pi(t ,\overrightarrow{x}) &=& \bar{\pi}(t) + \delta \pi(t ,\overrightarrow{x})\,\,.
\end{eqnarray} 
In the above $\bar{\xi}(t)$ ( where $\xi(t,\overrightarrow{x})=\rho(t,\overrightarrow{x})$, $p(t, \overrightarrow{x})$, ${u}^\mu$, ${\pi}(t,\overrightarrow{x})$) are the respective quantities for the homogeneous and isotropic background Universe and $\delta \xi(t,\overrightarrow{x})$ denotes their respective perturbations. Note that $v^\mu$ is perturbation of $u^\mu$ in Eq. (\ref{velocity}).

It is considered that perturbations are very small, therefore with $\delta T^\mu_\nu$ to be the perturbation for $T^\mu_\nu$ and using Eq. (\ref{energy_momentum}) one obtains, after neglecting second and higher order terms, 
\begin{eqnarray}
\delta T^0_0 & = & - \delta \rho \label{T00} \\
\delta T^0_i=-\delta T^i_0 & = & (\bar{\rho} + \bar{p})v^i\\
\delta T^i_j & = & \delta p \delta^i_j\,\,.\label{Tij}
\end{eqnarray} 
 For the Slotheon field tensor $T_{\mu\nu}^{(\pi)}$, the perturbations $\delta T^0_0$, $\delta T^0_i$, $\delta T^i_j$ are calculated using Eq. (\ref{slotheon_T}) and with Eqs. (\ref{T00} - \ref{Tij}) we now have,
\begin{equation}
\begin{split}
\delta\rho_\pi =-\frac{1}{a^3 M^2}\left[-a^3 M^2 \delta\pi V_\pi + 
   \dot{\pi} \left(2\dot{a} \nabla^2(\delta\pi) -\delta\dot{\pi}(a^3M^2 + 9 a \dot{a}^2) \right.\right.\\
   \left.\left.+ \dot{\pi} a \left(a^2 M^2 \Phi-\nabla^2 \Phi + 9\dot{a}(a\dot{\Phi}+2\Phi\dot{a})\right)\right)\right]
   \end{split}
 \end{equation}
\begin{equation}
   (\bar{\rho}_\pi+\bar{p}_\pi) v_i = -\frac{ \dot{\pi}}{a^2 M^2}\left[-2 a \delta \dot{\pi} \dot{a} + \delta \pi(a^2 M^2+ 3 \dot{a}^2)+a(a\dot{\Phi} + 3 \Phi \dot{a})\dot{\pi}\right]\mid_{i}
\end{equation}
\begin{equation}
\begin{split}
   \delta p_\pi = \frac{1}{a^3 M^2}\left[-a^3 M^2 \delta \pi V_\pi + \dot{\pi}\left(-2 a^2 \delta \ddot{\pi} \dot{a} + a (-a^2 M^2 \Phi + a^2 \ddot{\Phi}+\nabla^2 \Phi-\Phi_{ii})\dot{\pi} \right.\right.\\ 
   +\dot{a} (\nabla^2(\delta \pi)-\delta\pi_{ii}
   +2 a (3 a \dot{\Phi} + \Phi \dot{a})\dot{\pi})+ a \delta\dot{\pi}(-\dot{a}^2+a(a M^2-2\ddot{a}))\\ 
 \left.\left.   +4\Phi\dot{\pi}\ddot{a} a^2 \right)+a(\nabla^2\delta\pi-\delta\pi_{ii} - 2 a \delta\dot{\pi} \dot{a} + 2 a(a \dot{\Phi}+4 \Phi \dot{a})\dot{\pi})\ddot{\pi}\right]\,\,.
   \end{split}
\end{equation}
In the above $\delta \rho_\pi, v_i, \delta p_\pi$ are respectively the first ordered perturbed energy density, peculiar velocity and pressure density of the Slotheon field $\pi$. Also in the above, $\mid_i$ signifies the covariant derivative with 3 spatial coordinate, $\delta \pi_{ii}$ and $\Phi_{ii}$ are the double covariant derivatives of $\delta \pi$ and $\Phi$ respectively w.r.t the spatial co-ordinate $x_i$. 
\subsection{Linearised Perturbation Equations}

The perturbed Einstein's equation is given by
\begin{equation}
\delta G^\mu_\nu=8 \pi G \delta T^\mu_\nu \,\,, \label{pert EE}
\end{equation}
where $\delta G^\mu_\nu$ and $\delta T^\mu_\nu$ are perturbed Einstein's tensor and perturbed energy momentum tensor respectively.
Solving Eq. (\ref{pert EE}) for the perturbed space time metric (Eq. (\ref{metric})) we obtain the linearised Einstein's equations. The equations can now be written in the Fourier space by suitable Fourier decompositions of $\Phi, \delta \rho_i, \delta p_i$ etc and replacing $\nabla^2$ by $-k^2$. The equations then take the form \cite{DE}
\begin{eqnarray}
3 H^2 \Phi + 3 H \dot{\Phi} + \frac{k^2 \Phi}{a^2} &=&-4 \pi G \sum_i \delta \rho_i \label{pert EE1} \\
k^2(\dot{\Phi} + H \Phi) &=&  4 \pi G a\sum_i(\bar{\rho}_i+\bar{p}_i)\theta_i \label{pert EE2}\\
\ddot{\Phi} + 4 H \dot{\Phi} +2\dot{H} \Phi + 3 H^2 \Phi &=& 4 \pi G \sum_i \delta p_i\,\,. \label{pert EE3}
\end{eqnarray} 
It is to be mentioned here that for simplicity we have kept the notations of the quantities $\Phi, \delta \rho_i, \delta p_i$ etc unchanged while writing them in the Fourier space. Here the summation over $i$ represents the summation of the perturbations of the matter component (both dark matter and baryonic matter) and the perturbation of the Slotheon field. In what follows matter perturbations are designated by subscript $m$ and perturbations for Slotheon field are designated by subscript $\pi$. Since dust like particles have negligible pressure fluctuations, only Slotheon field contributes to the pressure density perturbations. The velocity gradient $\theta$ in the above takes the form $\theta=i \overrightarrow{k}\cdot\overrightarrow{v}$ in the Fourier space, $k$ being the wave number defined as $k=\frac{2 \pi}{\lambda_p}$ with $\lambda_p$  being the length scale of the perturbations. All perturbed quantities in the above equations correspond to the perturbations of the $k$th mode.
The dynamical equation for $\delta \pi$ can be obtained from the Slotheon action (Eq. (\ref{action})) for the perturbed space time metric (Eq. (\ref{metric})),
\begin{eqnarray}
\frac{1}{M^2}\left[H\dot{\pi} \frac{k^2 \Phi}{a} +18 \Phi \dot{a}^3\dot{\pi} - 2 \ddot{a} k^2 \delta {\pi} +2 \dot{a} \dot{\pi} k^2 \Phi -3\dot{a}^3 \delta \dot{\pi} + H^2 a (-k^2 \delta \pi)+\frac{\dot{\pi} k^2 \dot{\Phi}}{a} - a^3 M^2 V_{\pi\pi} \delta \pi + \nonumber \right.\\
2 M^2 a^3 \Phi(V_\pi+2\ddot{\pi}) +4 a^3 M^2 \dot{\pi} \dot{\Phi} + \dot{\pi} a^3 k^2\dot{\Phi} -a^3 M^2 \delta\ddot{\pi} + 36 a^2 \Phi \dot{\pi} H \ddot{a} + 18 \Phi a^3 H^2 \ddot{\pi} - M^2 a k^2 \delta \pi \nonumber\\  
+2 \ddot{\pi} a k^2 \Phi-6 a^2 H \ddot{a} \delta \dot{\pi} + 30 a^3 H^2 \dot{\pi}\dot{\Phi} +2 \dot{\pi} a k^2 \dot{\Phi} -3 a^3 H^2 \delta \ddot{\pi}  + 12 a^3 M^2 \Phi H \dot{\pi} + 6 \dot{\pi} a^2 \ddot{a} \dot{\Phi} \nonumber\\
 \left. -a^3 H(3 M^2 \delta \dot{\pi} + 6 \ddot{\pi} \dot{\Phi} - \dot{\pi} k^2 \Phi - 6 \dot{\pi} \ddot{\Phi})\right]=0 \,\,. \label{EOM del pi}
\end{eqnarray}
In the above $V_{\pi\pi}$ is the double derivative of the potential $V(\pi)$ w.r.t. $\pi$. 
Defining fractional density perturbation as
\begin{equation}
\delta = \frac{\delta \rho}{\bar{\rho}}\,\,.
\end{equation}
The quantity $\delta$ is computed for matter as well as Slotheon field by using the Eqs. (\ref{pert EE1} - \ref{EOM del pi}).
\subsection{Dimensionless Variables and Initial Conditions}

In order to solve the perturbation equations numerically one needs to define certain dimensionless variables and adopt certain well motivated initial conditions for these variables.

To this end the following dimensionless variables will be useful
for solving the background equations (Eq. (\ref{EE1} - \ref{EE3})) and the linearised perturbed equations (Eqs. (\ref{pert EE1} - \ref{EOM del pi})).
\begin{eqnarray}
x & = & \dfrac{\dot{\pi}}{\sqrt{6}H M_{{\rm pl}}} \label{x}\\
y & = & \dfrac{\sqrt{V(\pi)}}{\sqrt{3} H M_{{\rm pl}}}\label{y}\\
\lambda & = & -M_{{\rm pl}} \dfrac{V_\pi}{V(\pi)}\label{l}\\
\epsilon & = & \dfrac{H^2}{2M^2} \label{e}\\
q &=& \frac{\delta \pi}{\dfrac{d\pi}{dN}}\,\,.\label{q}
\end{eqnarray}
In the above $N={\rm ln}(a)$ is the number of e-foldings. Using these dimensionless variables in the Eqs. (\ref{EE1} - \ref{EE3}) and Eqs. (\ref{pert EE1} - \ref{EOM del pi}) the following autonomous system of equations are constructed
\begin{eqnarray}
\dfrac{dx}{dN} &=& \frac{P}{\sqrt{6}}-x \frac{\dot{H}}{H^2}\label{auto1}\\
\dfrac{dy}{dN} & = & -y \left(\sqrt{\dfrac{3}{2}} \lambda x + \dfrac{\dot{H}}{H^2}\right)\\
\dfrac{d\epsilon}{dN} & = & 2 \epsilon \dfrac{\dot{H}}{H^2}\\
\dfrac{d\lambda}{dN} & = & \sqrt{6} x \lambda^2 (1 - \Gamma)\\
\dfrac{dq}{dN} &=& q_1\label{autoq1}\\
\dfrac{d\Phi}{dN} &=& \Phi_1\label{autophi1}\\
\dfrac{dq_1}{dN} &=& \delta \pi_f-\frac{q \dot{H}}{x H^2} \dfrac{dx}{dN}- \frac{2 q_1}{x} \dfrac{dx}{dN} -q_1\frac{\dot{H}}{H^2}\\
\dfrac{d\Phi_1}{dN} &=& \frac{\delta p_f}{4\epsilon} -4 \Phi_1-2 \frac{\dot{H}}{H^2}\Phi-3\Phi-\Phi_1 \frac{\dot{H}}{H^2} \label{autolast}\,\,,
\end{eqnarray}
where $\Gamma=\frac{V V_{\pi\pi}}{V_\pi^2}$ and
\begin{eqnarray}
P &=&\frac{3 (12 \sqrt{6} x^3 \epsilon - 
   6 x^4 \beta \epsilon (1 + 18 \epsilon) + 
   y^2 \lambda + \beta (-1 + y^2) + 
   \sqrt{6} x (-1 -6 \epsilon y^2))}{1 + 6 \epsilon (1 + x^2 (-1 + 18 \epsilon))} \nonumber \\
   & & +\frac{3x^2 (-6 y^2 \epsilon \lambda + \beta (1 - 
      6 \epsilon (-4 + y^2)))}{1 + 6 \epsilon (1 + x^2 (-1 + 18 \epsilon))}\,\,,
      \end{eqnarray}
\begin{eqnarray}      
\frac{\dot{H}}{H^2} &=& \frac{12 \sqrt{6}
   x^3 \beta \epsilon (1 + 18 \epsilon) - 
 3 x^2 (1 + 6 \epsilon) (1 + 
    18 \epsilon) + (1 + 6 \epsilon) (-3 + 
    3 y^2)}{2+12\epsilon(1+x^2(-1+18\epsilon))}  \nonumber\\
    & & + 
 \frac{12 \sqrt{6}
   x \epsilon (y^2 \lambda + \beta (-1 + 
       y^2))}{2 + 12 \epsilon (1 + x^2 (-1 + 18  \epsilon))}\,\,,
\end{eqnarray}

In this work it is considered that dimensionless coupling constant $\beta=0$ (as mentioned before).
\begin{eqnarray}
\delta p_f = \frac{\delta p_\pi}{M^4} & = & \frac{6 \sqrt{6} \lambda  q x y^2 \epsilon -4 \epsilon ^2 \left(\frac{\dot{H}}{H^2} x+\frac{dx}{dN}\right) \left(6 x \left(3 L q -3 L_i q - 8 \Psi -2\Psi_1 \right)\right)}{1-\frac{6 x^2 \epsilon -108 x^2 \epsilon ^2}{6 \epsilon +1}}+ \nonumber \\
& &\frac{4 \epsilon ^2 \left(\frac{\dot{H}}{H^2} x+\frac{dx}{dN}\right)\left(12 (q \frac{dx}{dN}+ q_1 x)\right)}{1-\frac{6 x^2 \epsilon -108 x^2 \epsilon ^2}{6 \epsilon +1}}\nonumber \\
& & + \frac{12 x \epsilon  \left(24 \epsilon  \frac{d\epsilon}{dN}+(6 \epsilon +1) (2  (3-2 \frac{\dot{H}}{H^2}) \epsilon +1)\right)\left(q \frac{dx}{dN}+ q_1 x \right)}{6  \epsilon +1} \frac{1}{1-\frac{6 x^2 \epsilon -108 x^2 \epsilon ^2}{6 \epsilon +1}} \nonumber \\
& & -\frac{36 x \Psi  (2 \epsilon )^3 \left(a^2 L x+\frac{L x}{a^2}+2 (L \frac{dx}{dN}+3 x)\right)}{6  \epsilon +1}\frac{1}{1-\frac{6 x^2 \epsilon -108 x^2 \epsilon ^2}{6 \epsilon +1}}- \nonumber \\
& & \frac{2 x \epsilon }{6  \epsilon +1}\left(6 x \Psi  (2  \epsilon  (72 \epsilon +19)+1)-72  L q x \epsilon  (6   \epsilon +1)\right)\frac{1}{1-\frac{6 x^2 \epsilon -108 x^2 \epsilon ^2}{6 \epsilon +1}} \nonumber \\
& & +\frac{2 x \epsilon }{6 \alpha  \epsilon +1} \left(8 \sqrt{6}   \Psi  \epsilon  (2 \sqrt{6} \frac{dx}{dN} (9  \epsilon +1)-3 \lambda  y^2)\right)\frac{1}{1-\frac{6 x^2 \epsilon -108 x^2 \epsilon ^2}{6 \epsilon +1}} \nonumber \\
& & -\frac{12  L x \Psi_1 \epsilon }{a^2}-12 a^2   L x \Psi_1 \epsilon +3 q x \left(4  L \frac{d\epsilon}{dN}+(6  \epsilon +1) \left( L_i-L\right)\right)\frac{24 x \epsilon ^2}{6 \epsilon +1}\frac{1}{1-\frac{6 x^2 \epsilon -108 x^2 \epsilon ^2}{6 \epsilon +1}} \nonumber \\
& &+x \Psi  \left((6   \epsilon +1)\left(2  +4 \frac{\dot{H}}{H^2}-3 L+3 L_i+4 \right)-72   \frac{d\epsilon}{dN}\right)\frac{24 x \epsilon ^2}{6 \epsilon +1}\frac{1}{(1-\frac{6 x^2 \epsilon -108 x^2 \epsilon ^2}{6 \epsilon +1})} \nonumber \\
& & +2 \Psi_1 \left(x (3  (\epsilon  (6 -4 (L+7))-2 \frac{d\epsilon}{dN}+1)-4)+12   \frac{dx}{dN} \epsilon \right)\frac{24 x \epsilon ^2}{6 \epsilon +1}\frac{1}{(1-\frac{6 x^2 \epsilon -108 x^2 \epsilon ^2}{6 \epsilon +1})} \nonumber \\
& & -32 a \epsilon ^2 \left(2 (9  \Psi  \epsilon +\Psi )+3   (2 \epsilon ) (L \Psi -\Psi_1)\right)\frac{6 a x^2}{a^2 (6 \epsilon +1)}\frac{1}{(1-\frac{6 x^2 \epsilon -108 x^2 \epsilon ^2}{6 \epsilon +1})} \nonumber \\
& & +4 a \epsilon ^2 \left((6   \epsilon +1)-24   \epsilon \right)\left(-2 \frac{\dot{H}}{H^2} \Psi -3 \Psi -4 \Psi_1 \right)\frac{6 a x^2}{a^2 (6 \epsilon +1)}\frac{1}{(1-\frac{6 x^2 \epsilon -108 x^2 \epsilon ^2}{6 \epsilon +1})} \nonumber \\
& & +8 a q \epsilon ^2 \left(-\frac{3 y^2}{\sqrt{6} x}\frac{d\lambda}{dN}+3 y^2 \lambda^2 \right )\frac{6 a x^2}{a^2 (6 \epsilon +1)}\frac{1}{(1-\frac{6 x^2 \epsilon -108 x^2 \epsilon ^2}{6 \epsilon +1})} 
\end{eqnarray}
In the above $L=\frac{k^2}{3 a^2 H^2}$ and $ L_i=\frac{k_i^2}{3 a^2 H^2}$, where $k_i$ is the $i$th component ($i=x,y,z$) of $k$. Considering that the Universe is isotropic we assume for the present calculations that the $L_x=L_y=L_z=\frac{L}{3}$.
\begin{eqnarray}
\delta \pi_f = \frac{\delta \ddot{\pi}}{H^2 \dfrac{d\pi}{dN}} & = & \frac{6 \sqrt{6} \epsilon \left(L x \Phi + a^4 L x \Phi 6 a^2 x \Phi +2 a^2 L \dfrac{dx}{dN} \Phi \right) -3 a^2 \sqrt{6} L q x }{a^2 x \sqrt{6} (1 + 6 \epsilon)} \nonumber \\
& & +\frac{ -6 a^2 \sqrt{6} L q x(3 \epsilon + \dfrac{d\epsilon}{dN}) + a^2\left( 4 \sqrt{6} \dfrac{dx}{dN} \Phi (1+9\epsilon) + 4 \sqrt{6} x \Phi (2+9( \epsilon+ \dfrac{d\epsilon}{dN}))-6 y^2 \lambda \Phi\right)
}{a^2 x \sqrt{6} (1 + 6 \epsilon)} \nonumber \\
& & + 
\frac{-3 a^2 \sqrt{6}\left(1+2(3\epsilon + \dfrac{d\epsilon}{dN})\right)\left(q1 x+ q \dfrac{dx}{dN}\right) +\sqrt{6}\left(-6 x a^2 \Phi_1(\frac{-\dot{H}}{H^2} +1) 2\epsilon\right)}
{a^2 x \sqrt{6} (1 + 6 \epsilon)} \nonumber \\
& & + \frac{\sqrt{6}\left(6\Phi_1 L x \epsilon + 6 \Phi_1 a^4 L x \epsilon +2\Phi_1 a^2 \left(6\dfrac{dx}{dN} \epsilon +2 x + 6 x \epsilon (7+L)+ 3 x \dfrac{d\epsilon}{dN} \right)\right)}{a^2 x \sqrt{6} (1 + 6 \epsilon)} \nonumber \\
& & +\frac{\sqrt{6}\left(a^2 x \Phi (\frac{\dot{H}}{H^2} +1)(4+36 \epsilon+12 L\epsilon) +a^2 q x (\frac{3 y^2}{\sqrt{6}x}\dfrac{d\lambda}{dN}-3 \lambda y^2)\right)}{a^2 x \sqrt{6} (1 + 6 \epsilon)} \nonumber \\
& & + \frac{12\sqrt{6} a^2 x \epsilon \left(\frac{\delta p_f}{4\epsilon} - 4 \Phi_1 -2 \frac{\dot{H}}{H^2}\Phi-3\Phi \right)}{a^2 x \sqrt{6} (1 + 6 \epsilon)}\,\,,
\end{eqnarray}

\noindent {\underline {\it {The Initial Conditions}}}

The autonomous equations (Eqs. (\ref{auto1} - \ref{autolast})) are solved by adopting certain initial conditions for the background quantities as well as for the perturbed quantities. We choose the initial conditions at red shift $z\simeq1100$, i.e., at the early matter dominated Universe. We consider thawing Dark Energy models \cite{thawing}, \cite{thawing2} for the Slotheon field. In a thawing model, the equation of state (EOS) $\omega_\pi$ starts deviating from a frozen initial value of $-1$ with the progress of time. The initial conditions for thawing model indicate that $x_i$, the initial value of dimensionless quantity $x$ (Eq. (\ref{x})), is close to zero. In this case we take a very small value of $x_i$. The initial value $y_i$ of $y$ (Eq. (\ref{y})) is so chosen that the Dark Energy density parameter $\Omega_\pi$ and the matter density parameter $\Omega_m$ attain the values of around $0.70$ and $0.30$ respectively at the present epoch. The initial value $\lambda_i$ of $\lambda$ (Eq. (\ref{l})) that determines the slope of the potential $V(\pi)$ is adopted to be $0.7$. The initial value $\epsilon_i$ of $\epsilon$ (Eq. (\ref{e})) is treated as a parameter (it may be noted that the change in $\epsilon_i$ in fact indicates the change in the energy scale $M$). The variable $\epsilon$ contributes to the Slotheon term  $\dfrac{G^{\mu\nu}}{2M^2} \pi_{;\mu}\pi_{;\nu}$. 

It is observed that when the initial value $x_i \sim 0$ the results do not change significantly with the change of $x_i$, whereas the results are very sensitive to the choice of $y_i$. 
Noting that the dimensionless variable $\epsilon$ related to the Hubble parameter $H_0$ (as Eq. (\ref{e})), the initial values of $\epsilon$  are so chosen that at the present epoch $H_0$ attains a value of around $67.4$ km s$^{-1}$ Mpc$^{-1}$ \cite{planck}.

At a very early epoch of matter dominated Universe, there was no or negligible contribution of Dark Energy to the matter energy content of the Universe. In the present content, therefore Slotheon field had insignificant contribution at that era. In the present calculations we adopt small values for $q_i$ and $q_{1i}$, where at the initial epoch $q$ (Eq. (\ref{q}))$=q_i$ and $q_1 (q_1=\frac{dq}{dN},$ Eq (\ref{autoq1})$)=q_{1i}$. For choosing initial value of the gravitational potential we first write the Poisson's equation for gravitational potential (in Fourier space $\nabla^2=-k^2$)
\begin{equation}
k^2 \Phi = -4\pi G a^2 \sum_i \bar{\rho}_i \delta_i\,\,.
\end{equation}
Now in the early matter dominated epoch $\Omega_m=1$ and $\Omega_\pi=0$. Thus the initial condition of $\Phi$ can be obtained from the relation 
\begin{equation}
\Phi_i=-\frac{3}{2} \frac{H_i^2}{k^2} a_i^3\,\,,\label{poisson}
\end{equation}
where $\Phi_i$ is the initial gravitational potential. In Eq. (\ref{poisson}), it is considered that during matter dominated era matter density contrast $\delta_m$ is proportional to $a$. It is known that during the matter dominated era, $\Phi$ is almost constant. Thus at this epoch $\Phi_1 (\Phi_1=\frac{d\Phi}{dN}$, Eq. (\ref{autophi1}))$=\Phi_{1i}=0$.

\section{Numerical Solutions and Results}

The autonomous system of equations (Eqs. (\ref{auto1} - \ref{autolast})) are solved numerically using the initial conditions mentioned above and considering an exponential form for the potential $V(\pi)$ as given in Eq. (\ref{potential}).

\subsection{Equation of State and Density Parameters}

From the Einstein's equations of the background space time (Eqs. (\ref{EE1} - \ref{EE3})), density parameter $\Omega_\pi (=\frac{\bar{\rho}_\pi}{\rho_c})$ of the field $\pi$ and the matter density parameter $\Omega_m (=\frac{\bar{\rho}_m}{\rho_c})$, $\rho_c$ is the critical density of the Universe, are obtained as
\begin{eqnarray}
\Omega_\pi &=& y^2+x^2(1+18 \epsilon)\\
\Omega_m &=& 1-y^2-x^2(1+18 \epsilon)\,\,,
\end{eqnarray}
where $x, y$ etc. are defined in Eqs (28-32). The effective EOS ($\omega_{\rm eff}$) is derived as
\begin{equation}
\omega_{\rm eff} = \frac{p_{\rm total}}{\rho_{\rm total}} = -1-\frac{2 \dot{H}}{3 H^2}\,\,,
\end{equation}
where $ p_{\rm total}=p_{m} +p_\pi $and$ \rho_{\rm total} = \rho_{m} +\rho_\pi$. Hence for the flat FRW Universe, EOS $\omega_\pi$ of the Slotheon field $\pi$ takes the form 
\begin{equation}
\omega_\pi=\frac{\omega_{\rm eff}}{\Omega_\pi}\,\,.
\end{equation}
Using these equations, evolution of density parameters as with scale factor $a$ and the evolution of $\omega_\pi$ with redshift $z$ are calculated and the results are plotted in Fig. 1.
\begin{figure}[H]
\includegraphics[scale=0.5]{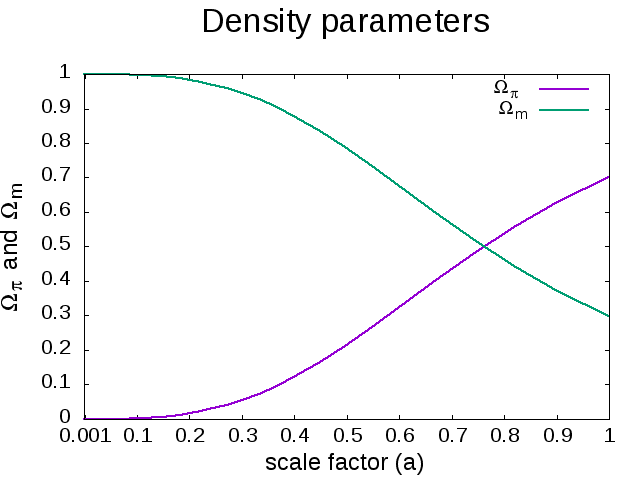}
\includegraphics[scale=0.5]{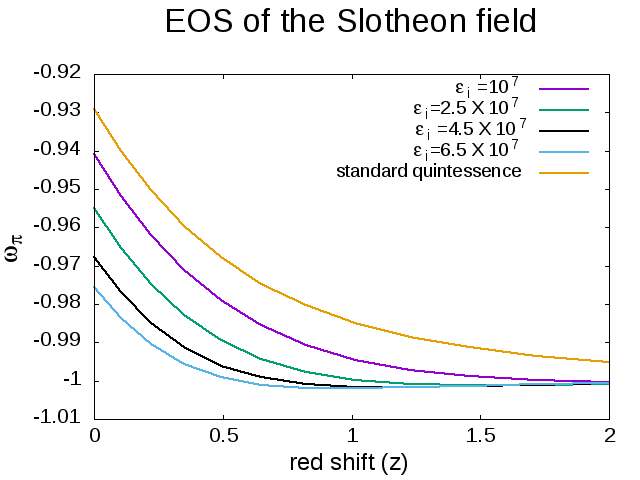}
\caption{(a) Variation of density parameters with scale factor $a$. (b)  Variation of EOS of the Slotheon field with redshift $z$.}
\end{figure}
From Fig. 1(a) it is seen that $\Omega_m$ is equal to $1$ and $\Omega_\pi$ is equal to zero at the early matter dominated  Universe. As the Universe undergoes evolution with time ($a$ increases) the matter density depletes where as the Dark Energy density (The Slotheon field density in the present work) grows. The cross over occurs at the epoch when $a \sim 0.77$ after which Dark Energy component starts dominating over the matter component of the Universe. This may be noted that the value of redshift $z$ at crossover point ($ a \sim 0.77$, Fig. 1(a)) is $z \sim 0.3$. Thus the phenomenon of Dark Energy domination and the consequent late time acceleration of the Universe happens at a recent cosmological past. Also from Fig. 1(a) this can be seen that at the present epoch ($a=1$) $\Omega_\pi\simeq0.7$ and $\Omega_m\simeq0.3$.

In Fig. 1(b) the variations of the EOS $\omega_\pi$ for the Slotheon field $\pi$ with redshift $z$ are shown for four different initial values of $\epsilon$, $\epsilon_i=10^7, 2.5\times 10^7, 4.5\times10^7, 6.5 \times 10^7$. Similar variations for the standard quintessence field are also shown in the same figure for comparison. One can see from Fig. 1(b) that for higher values $\epsilon$ the nature of the plots tend to $\Lambda$CDM value $-1$ and move away from that of quintessence. This can be explained by the fact that  the Slotheon term causes an extra slow roll to the scalar field  $\pi$ and hence the scalar field with the Slotheon term has more affinity towards the $\Lambda$CDM than the canonical scalar field without this term. 

\subsection{Density Fluctuations of Matter Field and Slotheon Field}

We solve numerically the autonomous set of equations (Eqs. (\ref{auto1} - \ref{autolast})) and obtain the variations of the gravitational potential $\Phi$, the density perturbation $\delta_\pi$ of the Slotheon field and the matter density contrast $\delta_m$ with the scale factor and the results are shown in Figs. 2-4. For these calculations the adopted initial values of different quantities are discussed in sect 3.

In Fig. 2 the evolution of the gravitational potential $\Phi$ with the scale factor $a$ are plotted for the three initial conditions of $\epsilon$, namely $2.5\times10^7, 4.5\times 10^7, 6.5 \times 10^7$. Similar variations of $\Phi$ are also shown for the standard quintessence field for comparison. We also compute the evolution of gravitational potential $\Phi$ with $a$ for the $\Lambda$CDM model and the results are plotted in same figure (Fig. 2). From Fig. 2 it is observed that in the early matter dominated epoch $\Phi$ is constant but as the Dark Energy component begins to contribute significantly, the gravitational potential suffers depletion. Note that in the early epoch  when there is negligible contribution of the Dark Energy component both the Slotheon and cosmological constant model behave identically in terms of the variation of $\Phi$ with $a$. But this is not so in later time when the Dark Energy component gradually increases. This is also to mention that the size of the perturbation for all the calculations in this section is taken to be $5\times10^2$ Mpc.
\begin{figure}[H]
\begin{center}
\includegraphics[scale=0.6]{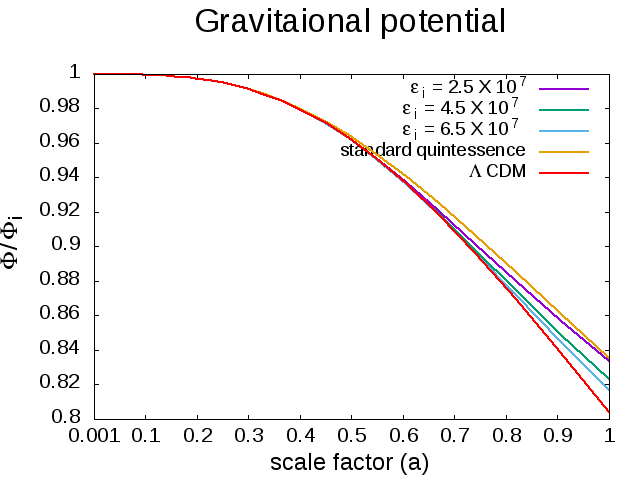}
\caption{Evolution of gravitational potential with scale factor for the Slotheon field}
\end{center}
\end{figure}

The density fluctuations $\delta_\pi$ and $\delta_m$ of Slotheon field $\pi$ and matter respectively are calculated by using the linearised Einstein's equations (Eqs. (\ref{pert EE1} - \ref{EOM del pi})) in terms of the dimensionless variables given in Eqs. (\ref{x} - \ref{q}) and they are given as follows
\begin{eqnarray}
\delta_\pi &=& \frac{1}{\Omega_\pi}\left(\sqrt{6} q \lambda y^2 x -24 q L \epsilon x^2 - 2 x^2 q_1 -2 q x \dfrac{dx}{dN}-36 x^2 \epsilon q_1  \right.\nonumber\\
& &  \left.- 36 \epsilon q x \dfrac{dx}{dN} + 2 x^2\Phi + 12 x^2 L \epsilon \Phi +72 x^2\epsilon\Phi + 36 x^2 \epsilon \dfrac{d\Phi}{dN}\right) \label{delpi} \,\,.
\end{eqnarray}
\begin{equation}
\delta_m =-\frac{1}{\Omega_m}\left(2 L \Phi + 2 \Phi + 2 \Phi_1 \right)-\delta_\pi\frac{\Omega_\pi}{\Omega_m}  \label{delm}
\end{equation}

In Fig. 3 the perturbations $\delta_\pi$ (perturbation of Dark Energy considered in this work) with $a$ are plotted for $\epsilon_i=2.5 \times 10^7, 4.5 \times 10^7, 6.5 \times 10^7$. As in Fig. 2 similar variations for standard quintessence field and $\Lambda$CDM  are also shown for comparison. It can be observed from Fig. 3 that at the early matter dominated epoch $\delta_\pi$ is zero, as expected. But with time (higher scale factor) the perturbation of Dark Energy deviates from zero and gradually increases with the rise in the contribution of Dark Energy. 

\begin{figure}[H]
\begin{center}
\includegraphics[scale=0.6]{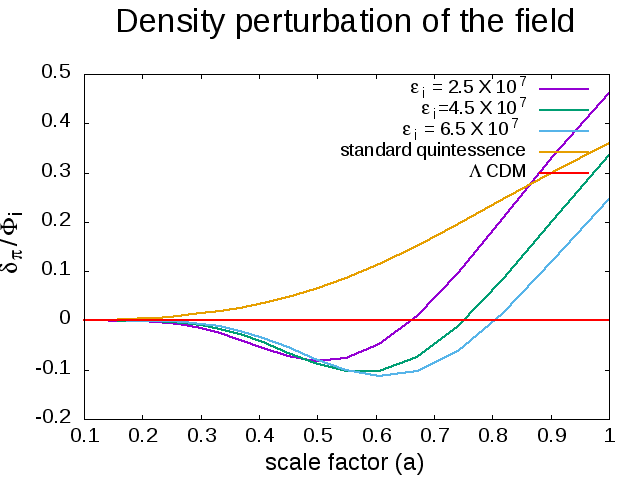}
\caption{Evolution of the density fluctuation of the Slotheon field with scale factor}
\end{center}
\end{figure}
 
Fig. 4 shows the evolution of matter density perturbation $\delta_m$ with the scale factor $a$ for the same initial values of $\epsilon$ adopted in Fig. 2 and 3. Quintessence and $\Lambda$CDM results are also shown for comparison. Here to at the initial stage $\delta_m$ is small and grows almost linearly in the matter dominated epoch, thus $\delta_m \sim a$ in the matter dominated era. This growth appears to be depleted to some extent in the Dark Energy dominated epoch ($a \gtrsim 0.77$). Note that $\delta_m$ in the Slotheon model coincides with that of $\Lambda$CDM and standard quintessence models in the early Universe but they deviate more with time. 
\begin{figure}[H]
\begin{center}
\includegraphics[scale=0.6]{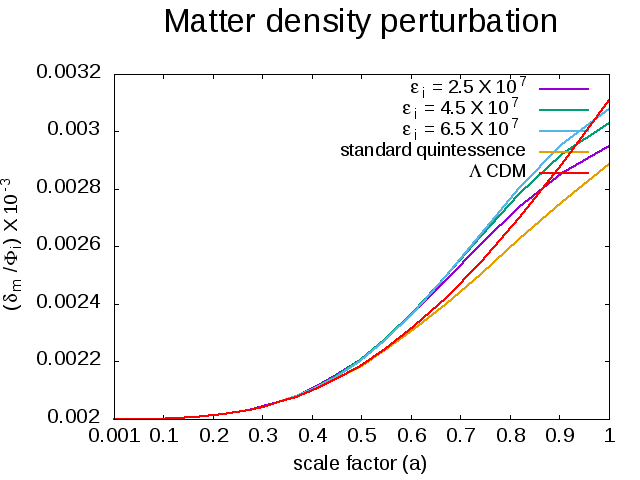}
\caption{Evolution of matter density fluctuations with scale factor for Slotheon field}
\end{center}
\end{figure}
  
\section{The Effect of the Slotheon Field on the Matter Power Spectrum }

In this section we explore the effect of the Slotheon Dark Energy perturbations on the matter power spectrum of the Universe. We compute the matter power spectrum with the Slotheon field and compare it with the same obtained from $\Lambda$CDM model. For $\Lambda$CDM model, cosmological constant $\Lambda$ is constant and there are no Dark Energy perturbations, but in the Slotheon model, with evolving equation of state $\omega_\pi$, the Dark Energy perturbations play important role in the nature of the power spectrum.

Matter power spectrum is defined as the average of the modulus square of the matter density fluctuation $\delta_m(k,a)$ and is given by \cite{DE}
\begin{equation}
Pm=\langle|\delta_m(k,a)|^2\rangle\,\,. 
\end{equation}

We compute the matter power spectrum $Pm_{\rm slotheon}$ for Slotheon field and that ($Pm_{\Lambda{\rm CDM}}$) for $\Lambda$CDM model using equations in section 3 and Eqs. (\ref{delpi} - \ref{delm}). We define a percentage suppression $X$ for the Slotheon power spectrum w.r.t. the power spectrum obtained from $\Lambda$CDM model as

\begin{equation}
\frac{Pm_{\Lambda {\rm CDM}}-Pm_{{\rm slotheon}}}{Pm_{\Lambda {\rm CDM}}} \times 100 = \frac{\Delta Pm}{Pm_{\Lambda {\rm CDM}}}\times 100=X\,\,.
\end{equation}

\begin{figure}[H]
\begin{center}
\includegraphics[scale=0.6]{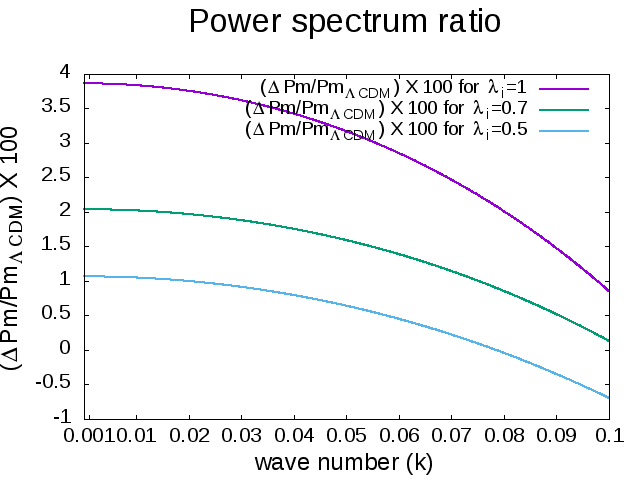}
\caption{Variations of the suppressions for the Slotheon power spectrum w.r.t. the $\Lambda$CDM  power spectrum, with wave number $k$}
\end{center}
\end{figure}

In Fig. 5 we plot the variations of $X$ with $k$ for $\epsilon_i=6.5 \times 10^7$ at $z=0$. Three such variations corresponding to three different fixed values of $\lambda_i$, namely $\lambda_i=0.5, 0.7, 1$, are shown in Fig. 5. From Fig. 5 it is seen that in general for lower values of $k$ the percentage suppressions $X$ is more which slowly diminishes as $k$ increases. For example for $\lambda_i=0.7$, the suppression $X \sim 2 {\text{\%}}$ for $k=10^{-3}$Mpc$^{-1}$ but $X$ is reduced to $1.8{\text{\%}}$ when $k=4 \times 10^{-2}$ Mpc$^{-1}$. This may be mentioned in the passing that for larger $k$ values (around $k \gtrsim 0.1h$ Mpc$^{-1}$) non linearity set seen \cite{nonlinear} and linear perturbation treatment may not be useful. From Fig. 5 it can also be noted that $X$ decreases with the decrease of $\lambda_i$. It is expected because $\lambda$ is related to the slope of the potential $V(\pi)$ (Eq. (\ref{potential})). Hence more $\lambda_i$ decreases, more flat the potential $V(\pi)$ tends to be and consequently approaches to $\Lambda$CDM model (which is based on a flat potential). It can also be mentioned from Fig. 5 that power spectrum for Slotheon field is not much different from $\Lambda$CDM power spectrum since the maximum suppression is only $\sim 3.8{\text{\%}}$ even for $\lambda_i=1$.

\section{Summary and Discussions}

The Dark Energy and late time acceleration of the Universe are addressed in this work by considering a type of scalar field, namely Slotheon scalar field, in a modified theory of gravity. Slotheon field is inspired by extra dimensional models at the dimensional cross over limit when Planck scale $M_{\rm pl} \rightarrow \infty$ and the theory is extended to curved space time. In order to address the inhomogeneities of Dark Energy and their evolutions, quantities such as matter density fluctuations, the perturbations of the scalar field etc. are worked out in this work. All the perturbation equations are then solved numerically.

The evolution of Dark Energy density $\Omega_\pi$ for the Slotheon scalar field $\pi$ in the potential $V(\pi)$ and the matter density $\Omega_m$ are calculated. The epoch of transition from matter dominated phase of the Universe to Dark Energy dominated phase (cross over epoch) is found to be at $z \simeq 0.3$ for the Slotheon Dark Energy considered here. The evolution of Dark Energy equation of state ($\omega_\pi$) is also calculated. The Dark Energy in the Slotheon field is considered to be a thawing type Dark Energy.

The perturbation equations for the Slotheon field and matter are then derived with a suitable metric and from Einstein's equations and they are solved numerically. This is done to study the evolution of Dark Energy density perturbations for the present case of Slotheon scalar field Dark Energy as well as evolution of matter perturbations in the same framework. These results are then compared with those for standard quintessence Dark Energy model of a scalar field and $\Lambda$CDM model for which the Dark Energy has a constant magnitude $\Lambda$.

A crucial component for all the studies and calculations is to choose proper initial conditions for the dimensionless variables. In the present work all the choices of such initial conditions are justified with proper arguments. For evolution of gravitational potential and the evolution of matter perturbations, we find that although in the early Universe the Slotheon model results coincide with those of $\Lambda$CDM and general quintessence it deviate away in later time. The Dark Energy density fluctuations deviate from zero with time as the Dark Energy density grows in the Universe. From the calculations and analysis in the present work it appears that the behaviour of Dark Energy from Slotheon model is more akin to the results of $\Lambda$CDM model than that from general quintessence model.

The nonlinear power spectrum can be obtained by considering the second order perturbations. This is for posterity.


\vspace{1cm}
\noindent {\bf Acknowledgements}

One of the authors (U.M.) acknowledges the CSIR, grant (NO. 09/489(0106)/2017-EMR-I). U.M. also thanks A.A. Sen for some useful comments and suggestions.

\end{document}